\documentclass[reprint,
               superscriptaddress,
               nobibnotes,
               amsmath,
               amssymb,
               aps,
               hidelinks, 
               pre]{revtex4-2}

\usepackage{siunitx}
\usepackage[version=4]{mhchem}
\usepackage{xfrac}
\DeclareSIUnit{\angstrom}{\text{Å}}
\usepackage{bm}

\newcommand{\papertitle}{Numerical and analytical strategies to propagate correlated uncertainties through the discrete Fourier transform}
\renewcommand{\Re}{\operatorname{Re}}
\renewcommand{\Im}{\operatorname{Im}}

\usepackage{showyourwork}
\usepackage{hyperref}
\usepackage[capitalise]{cleveref}

\begin{document}

\title{\papertitle}

\author{Tuguldur Zolzaya}
  \affiliation{Department of Chemistry, University of Oxford, South Parks Road, Oxford OX1 3QZ, UK.}
\author{Andrew R. McCluskey}
\email{andrew.mccluskey@bristol.ac.uk}
  \affiliation{Centre for Computational Chemistry, School of Chemistry, University of Bristol, Cantock's Close, Bristol, BS8 1TS, UK.}
  \affiliation{Diamond Light Source, Harwell Campus, Didcot, OX11 0DE, UK.}

\begin{abstract}
    The Fourier transform underpins a range of comparisons between atomistic simulations and scattering measurements, yet uncertainty propagation through it is seldom discussed, and correlations between data points are usually ignored. 
    Here, we give a pedagogical account of uncertainty propagation through the discrete Fourier transform, including the full covariance matrix. 
    Two routes are described: a numerical scheme that samples from the multivariate normal distribution describing the data, and an analytical scheme that treats the transform as a change of basis. 
    We discuss the scaling behaviour of these methods and demonstrate their utility on problems that link simulation to scattering: the Faber--Ziman partial structure factors of liquid water, and the dynamic structure factor obtained from a rotational autocorrelation function.
\end{abstract}

\maketitle 

\section{Introduction}
\label{sec:intro}

The Fourier transform (FT) is a foundational mathematical tool with applications found across X-ray and neutron scattering. 
An FT connects the real-space arrangement of atoms to the reciprocal-space intensities recorded in a diffraction measurement~\cite{Paalman1963, Youngs2019} and, in the case of neutron spectroscopy, connects the motions of atoms and molecules to the energy-transfer spectra measured~\cite{Bee1988}. 
The elastic or inelastic scattering functions we measure are, to a good approximation, the FT of real-space and real-time correlation functions of the sample. 
The FT is the bridge between the quantities that are natural to compute or model and the quantities that are actually measured.

Atomistic simulation has established itself as an important tool for analysing diffraction and spectroscopic measurements~\cite{Coles2023, Legein2024, Armstrong2022}.
Molecular dynamics (MD) simulations, for example, have been used to construct detailed mathematical models of the static and dynamic behaviour of molecular systems, which have then been used to interpret experimental observations~\cite{Richardson2026}.
Simulations provide an estimate of the real-space structure and real-time dynamics of a system, while X-ray and neutron scattering measurements typically yield reciprocal-space and energy-resolved observables; hence, the FT is an essential operation that enables the comparison between the two. 
The analysis of MD simulations is achieved by sampling many configurations or trajectories, providing not just point estimates of structure and dynamics but estimates that carry meaningful, quantifiable uncertainty~\cite{McCluskey2024}. 

The uncertainty provided by an MD simulation is necessary to enable meaningful comparison with experiments and to understand whether differences arise from, for example, force-field parameterisation. 
To allow this comparison, we must propagate the uncertainty through the FT. 
Despite this need, the propagation of uncertainty through an FT is not widely discussed in the scattering or simulation literature.
Some discussion in the signal processing community can be found; this typically assumes that uncertainties are independent and rarely considers the role of covariance~\cite{ForniesMarquina1997}. 
However, covariance, or correlation, between estimates plays an important role in estimating dynamics from molecular dynamics simulations as successive frames of a trajectory are inherently correlated, and neglecting these correlations can substantially misrepresent the uncertainty in the transformed quantity~\cite{McCluskey2024}. 

In this work, we aim to provide a pedagogical overview of uncertainty propagation through the discrete FT (DFT) by outlining a Monte Carlo sampling approach and describing an analytical solution. 
We focus here on the DFT due to its importance in numerical computation and, therefore, simulation and experimental analysis. 
We compare the efficiencies of these two approaches and show their application to a pair of relevant problems in linking simulation with scattering: (i) computing the Faber-Ziman partial structure factors and (ii) computing the dynamic structure factor from a rotational autocorrelation function. 
We make these approaches available as an open source Python package \textsc{fourierror}~\cite{FourierrorGithub}.

\section{Discrete Fourier transform} 

The DFT has a straightforward mathematical formula: 
\begin{equation}
    \mathcal{F}_k = \frac{2}{N} \sum_{n=0}^{N-1} x_n\,\exp{\left(-i2\pi\frac{k}{N}n\right)},
    \label{eqn:dft}
\end{equation}
and it transforms a sequence of $N$ complex numbers, $\left\{\mathbf{x}_n\right\} := x_0, x_1, \ldots, x_{N-1}$, to another sequence of the same length, $\left\{\bm{\mathcal{F}}_k\right\} := \mathcal{F}_0, \mathcal{F}_1, \ldots, \mathcal{F}_{N-1}$. 
This relation can be written as a linear transformation, 
\begin{equation}
    \bm{\mathcal{F}} = \frac{2}{N} \mathbf{W}_N\,\mathbf{x},
    \label{eqn:lt}
\end{equation}
where $\mathbf{W}_N$ is the so-called DFT matrix,
\begin{equation}
    \mathbf{W}_N = 
    \begin{bmatrix}
        1 & 1 & 1 & \cdots & 1 \\
        1 & \zeta & \zeta^2 & \cdots & \zeta^{N-1} \\
        1 & \zeta^2 & \zeta^4 & \cdots & \zeta^{2(N-1)} \\
        1 & \zeta^3 & \zeta^6 & \cdots & \zeta^{3(N-1)} \\
        \vdots & \vdots & \vdots & \ddots & \vdots \\
        1 & \zeta^{N-1} & \zeta^{2(N-1)} & \cdots & \zeta^{(N-1)(N-1)} \\
    \end{bmatrix},
    \label{eqn:mat}
\end{equation}
and $\zeta = \exp{\left(\sfrac{-i2\pi}{N}\right)}$.
As we will see, the ability to rewrite the DFT as a linear transformation is essential to produce the analytical solution for uncertainty propagation. 

A naïve implementation of the DFT given in \cref{eqn:dft,eqn:lt} scales computationally as $\mathcal{O}(N^2)$, though this can be improved by factorising the DFT matrix into a product of sparse factors. 
This fast Fourier transform (FFT) algorithm reduces the computational complexity to $\mathcal{O}(N\log{N})$. 
The FFT has seen numerous significant applications across scattering analysis~\cite{TenEyck1973, Kneller1995}.

The most common FFT algorithm is that from Cooley and Tukey~\cite{Cooley1965}, though it is believed to have been originally developed by Carl Friedrich Gauss~\cite{Heideman1984}. 
For simplicity, we outline here the radix-2 decimation-in-time form of the Cooley-Tukey algorithm, which assumes that $N$ is a power of two, though other implementations follow similar, but more complex, logic. 
The radix-2 form expresses the DFT of length $N$ recursively in terms of two DFTs each of size $\sfrac{N}{2}$ and rewrites the dense $\mathbf{W}_N$ matrix as a product of sparse matrices. 
This algorithm splits $\mathbf{x}$ into the even and odd indexed entries, i.e., $\left\{\mathbf{x}^{\textrm{even}}_{n}\right\} := x_0, x_2, \ldots, x_{N-2}$ and $\left\{\mathbf{x}^{\textrm{odd}}_{n}\right\} := x_1, x_3, \ldots, x_{N-1}$, and the DFT is then, 
\begin{widetext}
    \begin{equation}
        \mathcal{F}_k = \frac{2}{N}\left[\left(\mathbf{W}_{\sfrac{N}{2}}\,\mathbf{x}^{\textrm{even}}\right)_k + \exp{\left(-i2\pi\frac{1}{N}k\right)} \left(\mathbf{W}_{\sfrac{N}{2}}\,\mathbf{x}^{\textrm{odd}}\right)_k\right],\;\;\;\textrm{for}\;k=0,\ldots,\frac{N}{2}-1,
    \label{eqn:fft}
    \end{equation}
    \begin{equation}
        \mathcal{F}_{k + \sfrac{N}{2}} = \frac{2}{N}\left[\left(\mathbf{W}_{\sfrac{N}{2}}\,\mathbf{x}^{\textrm{even}}\right)_k - \exp{\left(-i2\pi\frac{1}{N}k\right)} \left(\mathbf{W}_{\sfrac{N}{2}}\,\mathbf{x}^{\textrm{odd}}\right)_k\right],\;\;\;\textrm{for}\;k=0,\ldots,\frac{N}{2}-1.
    \label{eqn:fft2}
    \end{equation}
\end{widetext}
The computational efficiency of the Cooley-Tukey algorithm stems from the fact that the terms between the $\left(\ldots\right)_k$ parentheses are computed once and reused in both \cref{eqn:fft,eqn:fft2}. 
Together \cref{eqn:fft,eqn:fft2} take two inputs and produce two outputs, differing only in the sign of the exponentially weighted term, and are termed a \emph{butterfly}, due to the shape of the data-flow diagram~\cite{Gentleman1966}. 
One stage of the algorithm applies $\sfrac{N}{2}$ butterflies across the array, and $\log_2 N$ stages complete the transform. 
Since a butterfly reads exactly two of its inputs, each stage is a sparse linear map. 

The index $k$ corresponds to the angular frequency
\begin{equation}
    \omega_k = \frac{2\pi k}{N\Delta},
    \label{eqn:freq}
\end{equation}
where $\Delta$ is the spacing of the input, giving a resolution of $\sfrac{2\pi}{N\Delta}$ and a maximum frequency of $\sfrac{\pi}{\Delta}$. 
By convention, indices $k \geq \sfrac{N}{2}$ are taken as the negative frequencies, $\omega_{k-N}$. 
For a real input, $\mathcal{F}(-\omega)$ is the complex conjugate of $\mathcal{F}(\omega)$, so the real part is even in $\omega$ and the imaginary part is odd. 
The prefactor of $\sfrac{2}{N}$ in \cref{eqn:dft} is chosen so that a sinusoid of amplitude $A$ gives peaks of magnitude $A$ at $\pm\omega$. 

\section{Numerical Approach}
\label{sec:num}

We start by outlining a numerical approach to propagate uncertainties through a DFT.
We typically assume that measured data may be generally described as a multivariate normal distribution, $\mathcal{N}\left(\boldsymbol{\mu}, \boldsymbol{\Sigma}\right)$, where $\boldsymbol{\mu}$ is a vector of length $N$ that defines the mean values for the data and $\boldsymbol{\Sigma}$ is the full covariance matrix for the data. 
The covariance matrix consists of diagonal elements representing variances and off-diagonal elements representing covariances, so for uncorrelated uncertainties, the covariance matrix is diagonal. 

By modelling the measured data with a multivariate normal distribution, random samples representing hypothetical repeats of our measurement can be drawn. 
To draw such a random sample, a matrix square root of the covariance matrix, for example from a Cholesky decomposition or an eigendecomposition, is used to find the matrix $\mathbf{L}$, such that, 
\begin{equation}
    \boldsymbol{\Sigma} = \mathbf{L}\mathbf{L}^\top.
\end{equation}
A vector $\mathbf{z}=\left(z_0, \ldots, z_{N-1}\right)$ is then constructed by drawing each $z_i$ independently from a standard normal distribution, $\mathcal{N}(0, 1)$. 
The vector $\mathbf{z}$ is then transformed by 
\begin{equation}
    \mathbf{x} = \boldsymbol{\mu} + \mathbf{L}\mathbf{z}
\end{equation}
to give the random sample.

\begin{figure*}
    \centering
    \includegraphics[width=0.8\textwidth]{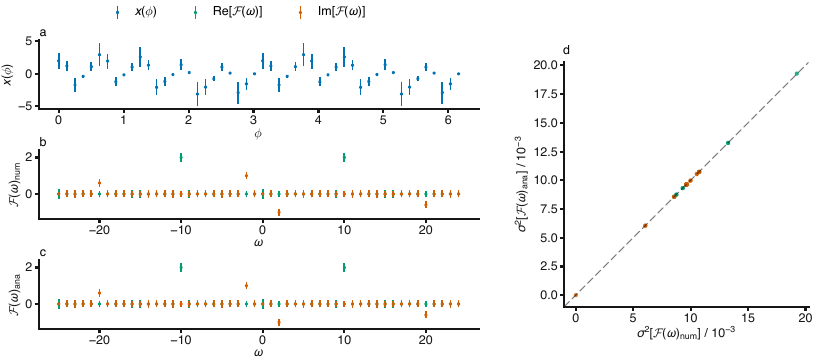}
    \caption{Comparison of the numerical and analytical propagation of uncertainty through a DFT (using the naïve algorithm): (a) the input data from \cref{eqn:input}, (b) the Fourier-transformed data with uncertainties determined numerically ($M = 10^6$), (c) the same, but with the uncertainty from the analytical solution, and (d) a comparison between the variances from the numerical and analytical approaches for the real (green) and imaginary (orange) parts.
    The error bars in (a)--(c) represent a \SI{95}{\percent} confidence interval, i.e., $1.96\sigma$.}
    \label{fig:compare}
    \script{compare.py}
\end{figure*}

By taking $M$ random samples from our measured data multivariate normal distribution, $\mathcal{N}_\mathbf{x}\left(\boldsymbol{\mu}_{\mathbf{x}}, \boldsymbol{\Sigma}_{\mathbf{x}}\right)$, and computing the DFT of each sample, using either the naïve algorithm or an FFT, we obtain $M$ samples from the FT of the data.
For a normally distributed input, the FT is similarly normally distributed.
This allows the real and imaginary components of these samples to be described with a mean, $\boldsymbol{\mu}_{\bm{\mathcal{F}}}$, and covariance matrix, $\boldsymbol{\Sigma}_{\bm{\mathcal{F}}}$.
The diagonal of the covariance matrix then contains the variances of the Fourier-transformed signal. 
The FT, with uncertainty estimation, for the function 
\begin{equation}
    x = \sin(2 \phi) + 2 \cos(10 \phi) + 0.6 \sin(20 \phi),
    \label{eqn:input}
\end{equation}
for $\phi$ uniformly sampled at 50 values on $[0, 2\pi)$ and a \qty{30}{\percent} relative uncertainty, is shown in \cref{fig:compare}.

\begin{figure}
    \centering
    \includegraphics[width=\columnwidth]{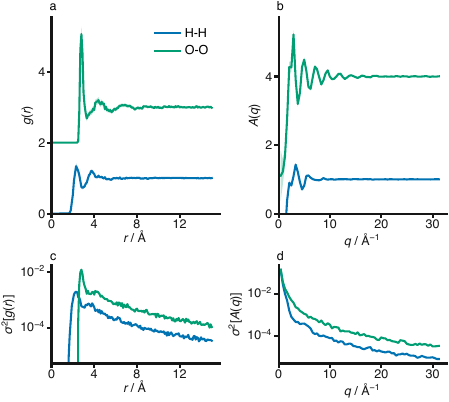}
    \caption{The H-H (blue) and O-O (green, offset for clarity in a and b) (a) partial pair distribution functions from a simulation of TIP4P water at \qty{298}{\kelvin} and (b) associated Faber-Ziman partial structure factors computed using an FT with uncertainties propagated using the numerical method ($M=5000$).
    The shaded regions indicate a \SI{95}{\percent} confidence interval, which is too narrow to be seen at this scale.
    The variances for the partial pair distribution functions and the Faber-Ziman partial structure factors are shown, on a logarithmic scale, in c and d, respectively.
    }
    \label{fig:water_pdf}
    \script{water_pdf.py}
\end{figure}

\cref{fig:water_pdf} shows the numerical uncertainty propagation applied to the determination of the Faber-Ziman partial structure factors, $A(q)$, from partial pair distribution functions, $g(r)$, taken from a simulation of liquid water. 
The Faber-Ziman partial structure factor is related to the sine transform, which is equivalent to the negative of the imaginary part of the FT, of $r\left[g\left(r\right) - 1\right]$~\cite{Faber1965}. 
How the uncertainty is propagated from the partial pair distribution function to the partial structure factors can be seen in \cref{fig:water_pdf}c \& d.
We highlight that the partial pair distribution functions are effectively uncorrelated, so we only present the variances in the main text. 
The full covariance matrices can be found in Appendix~\ref{app:full_covariance}.

To obtain the $g(r)$, a simulation of \num{3456} TIP4P water molecules was equilibrated at \qty{298}{\kelvin} under an NPT ensemble and then run for \qty{2.5}{\nano\second} with an NVT ensemble, using the final cell parameters from the NPT simulation as the starting configuration.
From the simulation, the partial pair distribution functions were found at \num{200} snapshots, evenly spaced across the trajectory, each computed from a different random selection of \qty{5}{\percent} of the molecules; the covariance matrix over these snapshots was then calculated. 

\section{Analytical Approach} 

\subsection{Naïve Algorithm}

As with \cref{sec:num}, the measured data are taken to be normally distributed, i.e., $\mathcal{N}_\mathbf{x}\left(\boldsymbol{\mu}_{\mathbf{x}}, \boldsymbol{\Sigma}_{\mathbf{x}}\right)$.
To perform the error propagation analytically, i.e., without using Monte Carlo sampling to estimate the variance, we note that for functions well described by linear transformations, error propagation can be performed by changing the basis of the covariance matrix. 
For some linear transformation, $\mathbf{f} = \mathbf{A}\mathbf{x}$, this change of basis is written, 
\begin{equation}
    \boldsymbol{\Sigma}_{\mathbf{f}} = \mathbf{A}\boldsymbol{\Sigma}\mathbf{A}^\top,
    \label{eqn:cov}
\end{equation}
where the subscript $\mathbf{f}$ indicates that $\boldsymbol{\Sigma}_{\mathbf{f}}$ is the covariance matrix of $\mathbf{f}$.

Euler's formula, $\exp{(i\theta)} = \cos(\theta) + i\sin(\theta)$, can be applied to each element of the DFT matrix, which allows us to write the matrix as real and imaginary parts: $\Re\left(\mathbf{W}_N\right) + i \Im\left(\mathbf{W}_N\right)$. 
Applying Euler's formula to an arbitrary matrix element, 
\begin{equation}
    \left(\mathbf{W}_N\right)_{kn} = \exp{-\left(i\frac{2{\pi}kn}{N}\right)}, 
\end{equation}
gives, 
\begin{equation}
    \Re \left[\left(\mathbf{W}_N\right)_{kn}\right] = \cos\left({\frac{2{\pi}kn}{N}}\right)
\end{equation}
and
\begin{equation}
    \Im \left[\left(\mathbf{W}_N\right)_{kn}\right] = -\sin\left({\frac{2{\pi}kn}{N}}\right).
\end{equation}
By rewriting \cref{eqn:dft} to use the DFT matrix,
\begin{equation}
    \mathcal{F}_k = \frac{2}{N} \sum_{n=0}^{N-1}{\left(\mathbf{W}_N\right)_{kn} x_n},
\end{equation}
Euler's formula can be applied to the entire matrix, 
\begin{widetext}
    \begin{equation}
        \begin{aligned}
        \mathcal{F}_k & = \Re\left(\mathcal{F}_k\right) + i\Im\left(\mathcal{F}_k\right) 
        = \frac{2}{N} \sum_{n=0}^{N-1}{\left\{\Re\left[\left(\mathbf{W}_N\right)_{kn}\right] +i\Im\left[\left(\mathbf{W}_N\right)_{kn}\right]\right\}x_n} 
        = \frac{2}{N} \sum_{n=0}^{N-1}{\left[\cos\left({\frac{2{\pi}kn}{N}}\right) - i\sin\left({\frac{2{\pi}kn}{N}}\right)\right]x_n} \\ 
        & = \left\{\frac{2}{N} \sum_{n=0}^{N-1}{\left[\cos\left({\frac{2{\pi}kn}{N}}\right)\right]x_n}\right\} - i\left\{\frac{2}{N} \sum_{n=0}^{N-1}{\left[\sin\left({\frac{2{\pi}kn}{N}}\right)\right]x_n}\right\}
        ,
        \end{aligned}
    \end{equation}
\end{widetext}
to show that the real and imaginary parts are individual linear transformations.  
This allows the real and imaginary parts of the DFT to be written individually as, 
\begin{equation}
    \Re(\bm{\mathcal{F}}) = \frac{2}{N}\Re\left(\mathbf{W}_N\right)\mathbf{x}
\end{equation}
and
\begin{equation}
    \Im(\bm{\mathcal{F}}) = \frac{2}{N}\Im\left(\mathbf{W}_N\right)\mathbf{x}.
\end{equation}
Therefore, to propagate the real and imaginary covariance matrices of the DFT result, $\boldsymbol{\Sigma}_{\Re\left(\bm{\mathcal{F}}\right)}$ and $\boldsymbol{\Sigma}_{\Im\left(\bm{\mathcal{F}}\right)}$, the general \cref{eqn:cov} can be used to give, 
\begin{equation}
    \boldsymbol{\Sigma}_{\Re\left(\bm{\mathcal{F}}\right)} = \left(\frac{2}{N}\right)^2 \Re\left(\mathbf{W}_N\right)\boldsymbol{\Sigma}_{\mathbf{x}} \Re\left(\mathbf{W}_N\right)^\top, 
    \label{eqn:real}
\end{equation}
and
\begin{equation}
    \boldsymbol{\Sigma}_{\Im\left(\bm{\mathcal{F}}\right)} = \left(\frac{2}{N}\right)^2 \Im\left(\mathbf{W}_N\right)\boldsymbol{\Sigma}_{\mathbf{x}} \Im\left(\mathbf{W}_N\right)^\top.
    \label{eqn:imag}
\end{equation}

\cref{fig:compare}(c) shows the result of this analytical procedure to propagate uncertainty, which is similar to the numerical result in \cref{fig:compare}(b). 
The variances from the two approaches are compared in \cref{fig:compare}(d), where it is clear that the variance values are similar, within numerical limits. 
The agreement between the variances from the numerical and analytical solutions improves as a function of $M$, as shown in Appendix~\ref{app:convergence}. 

\begin{figure}
    \centering
    \includegraphics[width=\columnwidth]{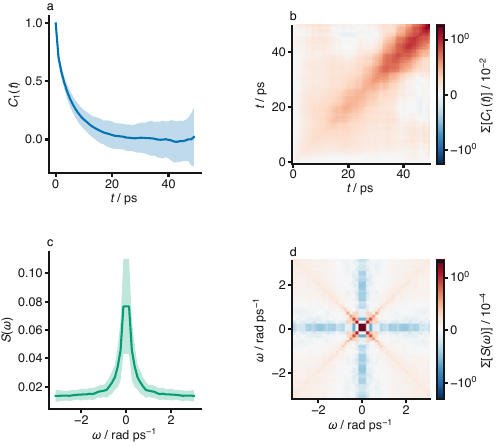}
    \caption{(a) The first-order H-H rotational autocorrelation function, $C_1(t)$, from the water simulation discussed previously, and (b) its full covariance matrix, obtained from sampling \num{50} sub-trajectories of \num{16} molecules, each \qty{50}{\pico\second} in length.
    (c) The quasi-elastic part of the dynamic structure factor, at $q=\qty{1.0}{\per\angstrom}$, obtained by analytically Fourier transforming the rotational intermediate scattering function, $F(q, t) = \sum_{\ell=0}^{2} (2\ell+1)\,j_\ell^2(qd)\,C_\ell(t)$, where $j_\ell$ is a spherical Bessel function and $d$ the H-H distance, and (d) the covariance matrix of the dynamic structure factor; the elastic contribution at $\omega=0$ is omitted.
    The shaded regions in (a) and (c) indicate a \qty{95}{\percent} confidence interval.}
    \label{fig:water_acf}
    \script{water_acf.py}
\end{figure}

To assess the analytical FT, we consider the correlated rotational autocorrelation function data from the same simulation of TIP4P water.
An FT of the rotational intermediate scattering function, constructed from the rotational autocorrelation functions, gives the dynamic structure factor, $S(\omega)$, at a given $q$-vector, which is itself a quantity that can be measured with neutron spectroscopy. 
\cref{fig:water_acf} shows the rotational autocorrelation function and the non-diagonal covariance matrix, alongside the resulting dynamic structure factor and propagated covariance matrix of $S(\omega)$. 
The full input covariance matrix, including the correlation between time points, is propagated, and this correlation leads to appreciable off-diagonal covariance in $S(\omega)$. 
The symmetry of the output covariance matrix about both diagonals is a consequence of the input being real, so that $\Re\left[\mathcal{F}(\omega)\right] = \Re\left[\mathcal{F}(-\omega)\right]$, rather than of the correlation. 

\subsection{Fast Fourier Transform}

The analytical result of \cref{eqn:real,eqn:imag} is a general statement about a linear transformation, and is therefore equally applicable to the FFT. 
The Cooley-Tukey factorisation does not change the linear map, $\mathbf{x} \mapsto \bm{\mathcal{F}}$; it only reorganises the evaluation. 
As a result, the covariance matrices of the real and imaginary parts of an FFT can be computed in the same way as for a DFT.  
However, to achieve the computational benefit of the FFT, the FFT factorisation should be used in the uncertainty propagation. 

The butterfly structure introduced above may be written explicitly as a matrix factorisation~\cite{VanLoan1992}. 
Collecting the $\sfrac{N}{2}$ butterflies of each level into a single stage matrix $\mathbf{P}_l$, and writing the reordering that precedes them as a bit-reversal permutation $\mathbf{B}$,
\begin{equation}
    \mathbf{W}_N = \mathbf{P}_{\log_2 N} \cdots \mathbf{P}_2
                   \mathbf{P}_1 \mathbf{B},
    \label{eq:ct_factorisation}
\end{equation}
in which each $\mathbf{P}_l$ inherits the sparsity of the butterflies it contains, with exactly two non-zero entries per row.

\cref{eqn:cov} can be used to carry the input uncertainty through the sequence of stages. 
As the stage matrices are complex, two second-moment matrices must be carried: the covariance, $\boldsymbol{\Gamma}$, and the pseudo-covariance, $\boldsymbol{\Pi}$, 
\begin{equation}
    \boldsymbol{\Gamma}^{(0)} = \boldsymbol{\Pi}^{(0)} = \mathbf{B} \boldsymbol{\Sigma}_{\mathbf{x}} \mathbf{B}^{\top},
    \label{eqn:stagewise2}
\end{equation}
\begin{equation}
    \begin{aligned}
    \boldsymbol{\Gamma}^{(l)} = \mathbf{P}_l \, \boldsymbol{\Gamma}^{(l-1)} \, \mathbf{P}_l^{\dagger}, & \;\;\;
    \boldsymbol{\Pi}^{(l)} = \mathbf{P}_l \, \boldsymbol{\Pi}^{(l-1)} \, \mathbf{P}_l^{\top}, \\
    \textrm{for}\;l & =1,\ldots,\log_2 N,
    \end{aligned}
    \label{eqn:stagewise}
\end{equation}
where $\dagger$ denotes the conjugate transpose. 
After the final stage, $\boldsymbol{\Gamma} = \mathbf{W}_N \boldsymbol{\Sigma}_{\mathbf{x}} \mathbf{W}_N^{\dagger}$ and $\boldsymbol{\Pi} = \mathbf{W}_N \boldsymbol{\Sigma}_{\mathbf{x}} \mathbf{W}_N^{\top}$, and the covariance matrices of the real and imaginary parts follow from Euler's formula as
\begin{equation}
    \boldsymbol{\Sigma}_{\Re\left(\bm{\mathcal{F}}\right)} = \left(\frac{2}{N}\right)^2 \frac{\Re\left(\boldsymbol{\Gamma}\right) + \Re\left(\boldsymbol{\Pi}\right)}{2},
    \label{eqn:stagewise_reim}
\end{equation}
and
\begin{equation}
    \boldsymbol{\Sigma}_{\Im\left(\bm{\mathcal{F}}\right)} = \left(\frac{2}{N}\right)^2 \frac{\Re\left(\boldsymbol{\Gamma}\right) - \Re\left(\boldsymbol{\Pi}\right)}{2}.
    \label{eqn:stagewise_reim2}
\end{equation}
Neither $\boldsymbol{\Gamma}$ nor $\boldsymbol{\Pi}$ alone is sufficient to recover both of these. 
Each stage requires two sparse multiplications, one from each side, for each of $\boldsymbol{\Gamma}$ and $\boldsymbol{\Pi}$, at a cost of $\mathcal{O}\left(N^2\right)$ per stage; over $\log_2 N$ stages the total is $\mathcal{O}\left(N^2 \log N\right)$.
Appendix~\ref{app:fft_implementation} outlines some specific implementation details for this approach.

\section{Computational Complexity}

\begin{figure*}
    \centering
    \includegraphics[width=\textwidth]{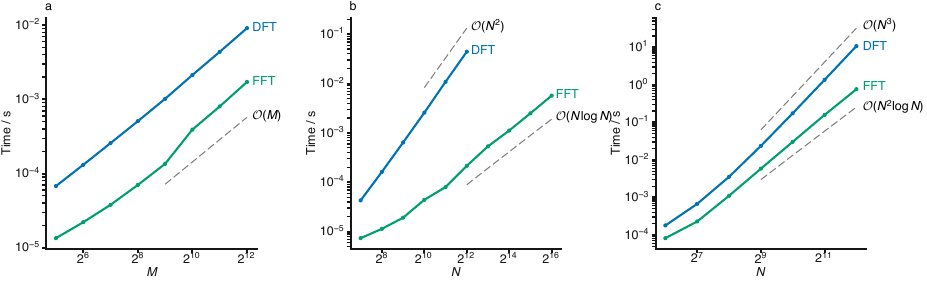}
    \caption{Computational complexity of the numerical method with (a) the number of samples, for a dataset with \num{100} points, and (b) the number of data points, for $M=10$; and (c) of the analytical method with the number of data points.
    For the numerical method, only the transforms are timed; drawing the samples is excluded.
    Evaluations were run on an Apple MacBook Pro with M2 Pro chipset and 16 GB RAM.
    The minimum time is shown, from \num{7} repeats in (a) and (b), and in (c) from \num{20} repeats for $N\leq512$, \num{5} for $512<N\leq2048$, and \num{3} for $N>2048$.}
    \label{fig:scaling}
    \script{scaling.py}
\end{figure*}

The numerical approach has relatively straightforward computational complexity, as performing $M$ individual DFTs or FFTs on samples from a multivariate normal distribution simply multiplies the cost of the underlying algorithm by $M$. 
Therefore, for a DFT the scaling is anticipated to be $\mathcal{O}\left(MN^2\right)$, where $N$ is the number of data points, while for an FFT the scaling will be $\mathcal{O}\left(MN\log N\right)$. 
These scalings count only the transforms; for correlated data, drawing the samples additionally costs $\mathcal{O}\left(N^3\right)$ for the matrix square root and $\mathcal{O}\left(MN^2\right)$ for the products $\mathbf{L}\mathbf{z}$, which can dominate. 
The computational complexity of the transforms alone is shown in \cref{fig:scaling}a and b. 
For the naïve analytical approach, we recognise that the DFT itself is no longer the computational bottleneck. 
The two successive matrix multiplications in each of \cref{eqn:real,eqn:imag} result in an overall complexity of $\mathcal{O}\left(N^3\right)$, which will dominate over the $\mathcal{O}\left(N^2\right)$ of the DFT. 
This cubic scaling behaviour is shown in \cref{fig:scaling}c alongside the expected analytical FFT $\mathcal{O}\left(N^2\log{N}\right)$ scaling.

\section{Conclusions}

We outline two approaches to propagate uncertainty, including any covariance, through a DFT.
The numerical approach draws samples from a multivariate normal distribution that describes the data, transforms each, before finding the covariance matrix of the results. 
Its statistical error decreases as $\sfrac{1}{\sqrt{M}}$ and its cost is that of $M$ transforms, which an FFT reduces from $\mathcal{O}(M N^2)$ to $\mathcal{O}(M N \log N)$. 
The analytical route instead exploits the linearity of the transform, propagating the covariance exactly by a change of basis, \cref{eqn:cov}. 
Written directly this costs $\mathcal{O}(N^3)$, dominated by the matrix products of \cref{eqn:real,eqn:imag} rather than by the transform itself; carrying the covariance through the butterfly stages of the Cooley--Tukey factorisation reduces this to $\mathcal{O}(N^2 \log N)$.

The two approaches allow a trade-off to be made; where $N\gg M$, it may be computationally desirable to use the less accurate numerical approach, whereas when accuracy is paramount, the analytical solution is available. 
Both routes are available in the \textsc{fourierror}~\cite{FourierrorGithub} Python package. 

\section*{Author Contributions}

T.Z.: Formal analysis, Funding acquisition, Investigation, Methodology, Software, Visualization, Writing – original draft.
A.R.M.: Conceptualization, Formal analysis, Investigation, Methodology, Project administration, Software, Supervision, Validation, Visualization, Writing – original draft, Writing – reviewing \& editing. \\

\section*{Acknowledgements}

T.Z. would like to acknowledge the Crankstart Internship Programme at the University of Oxford for funding his research stay at the University of Bristol.
The authors used Claude Sonnet 5.5 (Anthropic, accessed Sept 2026) during manuscript preparation.
Claude supported three tasks: drafting summaries of related work from relevant publications, suggesting revisions for clarity, and generating draft Python code for analysis.
The authors verified all citations against the original sources, rewrote the AI-assisted text, inspected and tested the code, and made all analytical and interpretive decisions.
Claude did not select the research question, collect data, determine the conclusions, or approve the final manuscript.

\section*{Data Availability and Reproducibility Statement}

The molecular dynamics simulation data that support the findings of this study are openly available on Zenodo at \href{https://doi.org/10.5281/zenodo.24078901}{10.5281/zenodo.24078901}.
A complete set of analysis/plotting scripts, allowing for a fully reproducible and automated analysis workflow using showyourwork, is available in a GitHub repository~\cite{Zolzaya2026} under an MIT license.
All of the methods shown here are available in the open source Python package \textsc{fourierror}~\cite{FourierrorGithub}.\\

\appendix

\section{Full Covariance Matrices of Partial Pair Distribution Functions and Structure Factors}
\label{app:full_covariance}

\begin{figure*}
    \centering
    \includegraphics[width=\textwidth]{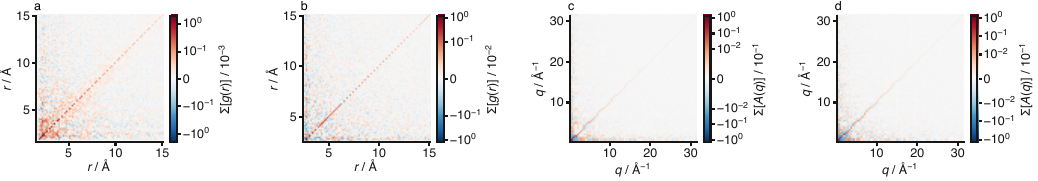}
    \caption{Full covariance matrices for the partial pair distribution functions and Faber--Ziman partial structure factors of TIP4P water.
    (a) and (b) show $\Sigma[g(r)]$ for the H--H and O--O partials,
    respectively, and (c) and (d) the corresponding $\Sigma[A(q)]$. 
    Note the differing colour scales; each colour scale is centred on zero and is linear close to zero and logarithmic beyond.
    Each $\Sigma[g(r)]$ is shown above the distance of closest approach, below which $g(r)$ vanishes identically, and is dominated by its diagonal, with the largest variances at the first coordination shell; the off-diagonal elements are consistent with zero.
    The structure factor covariances carry appreciable off-diagonal weight by contrast, showing that the FT introduces correlation between $q$ points even where the real-space input is uncorrelated.}
    \label{fig:water_cov}
    \script{water_cov.py}
\end{figure*}

\cref{fig:water_cov} shows the near-diagonal covariance matrices for the H-H and O-O pair distribution functions and the resulting, markedly less diagonal, covariance matrices for the H-H and O-O Faber-Ziman partial structure factors. 

\section{Convergence of the Numerical Approach}
\label{app:convergence}

\begin{figure}
    \centering
    \includegraphics[width=\columnwidth]{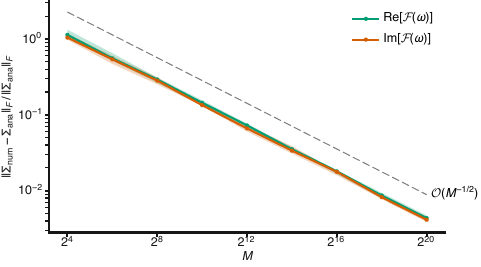}
    \caption{Relative error in the numerically estimated covariance matrices of the real (green) and imaginary (orange) parts of the FT of \cref{eqn:input}, as a function of the number of samples, $M$.
    The lines show the median and the shaded regions the central \SI{95}{\percent} of \num{10} repeats at each $M$.
    The dashed line indicates $\mathcal{O}\left(M^{-1/2}\right)$ scaling.}
    \label{fig:convergence}
    \script{convergence.py}
\end{figure}

The accuracy of the numerical approach is assessed by comparing the covariance matrices from $M$ samples with the exact analytical result, for the input in \cref{eqn:input}.
The error is measured as the relative Frobenius norm, $\left\|\boldsymbol{\Sigma}_{\mathrm{num}} - \boldsymbol{\Sigma}_{\mathrm{ana}}\right\|_F / \left\|\boldsymbol{\Sigma}_{\mathrm{ana}}\right\|_F$, of the real and imaginary covariance matrices.
As \cref{fig:convergence} shows, this error decreases as $M^{-1/2}$, as expected for a Monte Carlo estimate.
Therefore, each additional significant figure of accuracy requires a hundredfold increase in $M$.

\section{Analytical FFT Implementation Details}
\label{app:fft_implementation}

In practice \cref{eqn:stagewise,eqn:stagewise2} need not be implemented directly.
Since $(\mathbf{W}_N)_{kn} = \exp(-i 2 \pi k n / N)$ is symmetric, $\mathbf{W}_N^\top = \mathbf{W}_N$, and the two-sided product is a two-dimensional transform of the covariance matrix. 
Defining
\begin{equation}
    \mathbf{P} = \mathbf{W}_N \boldsymbol{\Sigma}_{\mathbf{x}} \mathbf{W}_N,
\end{equation}
\begin{equation}
    \mathbf{Q} = \mathbf{W}_N \boldsymbol{\Sigma}_{\mathbf{x}} \mathbf{W}_N^{*},
    \label{eq:PQ}
\end{equation}
where $\mathbf{P}$ is the two-dimensional DFT of $\boldsymbol{\Sigma}_{\mathbf{x}}$ and $\mathbf{Q}$ is obtained by transforming forward along one axis and backward along the other (i.e., $\mathbf{P} = \boldsymbol{\Pi}$ and $\mathbf{Q} = \boldsymbol{\Gamma}$ after the final stage of \cref{eqn:stagewise}), the real and imaginary covariance matrices follow from Euler's formula as
\begin{equation}
    \boldsymbol{\Sigma}_{\mathrm{Re}(\boldsymbol{\mathcal{F}})}
        = \left( \frac{2}{N} \right)^{\!2}
          \frac{\mathrm{Re}(\mathbf{P}) + \mathrm{Re}(\mathbf{Q})}{2},
    \label{eq:fft_sigma_re}
\end{equation}
\begin{equation}
    \boldsymbol{\Sigma}_{\mathrm{Im}(\boldsymbol{\mathcal{F}})}
        = \left( \frac{2}{N} \right)^{\!2}
          \frac{\mathrm{Re}(\mathbf{Q}) - \mathrm{Re}(\mathbf{P})}{2}.
    \label{eq:fft_sigma_im}
\end{equation}
These are algebraically identical to \cref{eqn:real,eqn:imag,eqn:stagewise_reim,eqn:stagewise_reim2}.
Each of $\mathbf{P}$ and $\mathbf{Q}$ costs $2N$ one-dimensional FFTs, giving the same $\mathcal{O}(N^2 \log N)$ scaling as \cref{eqn:stagewise,eqn:stagewise2} while reusing an existing FFT implementation.

\bibliography{bib}

\begin{thebibliography}{18}%
\makeatletter
\providecommand \@ifxundefined [1]{%
 \@ifx{#1\undefined}
}%
\providecommand \@ifnum [1]{%
 \ifnum #1\expandafter \@firstoftwo
 \else \expandafter \@secondoftwo
 \fi
}%
\providecommand \@ifx [1]{%
 \ifx #1\expandafter \@firstoftwo
 \else \expandafter \@secondoftwo
 \fi
}%
\providecommand \natexlab [1]{#1}%
\providecommand \enquote  [1]{``#1''}%
\providecommand \bibnamefont  [1]{#1}%
\providecommand \bibfnamefont [1]{#1}%
\providecommand \citenamefont [1]{#1}%
\providecommand \href@noop [0]{\@secondoftwo}%
\providecommand \href [0]{\begingroup \@sanitize@url \@href}%
\providecommand \@href[1]{\@@startlink{#1}\@@href}%
\providecommand \@@href[1]{\endgroup#1\@@endlink}%
\providecommand \@sanitize@url [0]{\catcode `\\12\catcode `\$12\catcode
  `\&12\catcode `\#12\catcode `\^12\catcode `\_12\catcode `\%12\relax}%
\providecommand \@@startlink[1]{}%
\providecommand \@@endlink[0]{}%
\providecommand \url  [0]{\begingroup\@sanitize@url \@url }%
\providecommand \@url [1]{\endgroup\@href {#1}{\urlprefix }}%
\providecommand \urlprefix  [0]{URL }%
\providecommand \Eprint [0]{\href }%
\providecommand \doibase [0]{https://doi.org/}%
\providecommand \selectlanguage [0]{\@gobble}%
\providecommand \bibinfo  [0]{\@secondoftwo}%
\providecommand \bibfield  [0]{\@secondoftwo}%
\providecommand \translation [1]{[#1]}%
\providecommand \BibitemOpen [0]{}%
\providecommand \bibitemStop [0]{}%
\providecommand \bibitemNoStop [0]{.\EOS\space}%
\providecommand \EOS [0]{\spacefactor3000\relax}%
\providecommand \BibitemShut  [1]{\csname bibitem#1\endcsname}%
\let\auto@bib@innerbib\@empty
%</preamble>
\bibitem [{\citenamefont {Paalman}\ and\ \citenamefont
  {Pings}(1963)}]{Paalman1963}%
  \BibitemOpen
  \bibfield  {author} {\bibinfo {author} {\bibfnamefont {H.~H.}\ \bibnamefont
  {Paalman}}\ and\ \bibinfo {author} {\bibfnamefont {C.~J.}\ \bibnamefont
  {Pings}},\ }\bibfield  {title} {\bibinfo {title} {{F}ourier {A}nalysis of
  {X}-{R}ay {D}iffraction {D}ata from {L}iquids},\ }\href
  {https://doi.org/10.1103/revmodphys.35.389} {\bibfield  {journal} {\bibinfo
  {journal} {Rev. Mod. Phys.}\ }\textbf {\bibinfo {volume} {35}},\ \bibinfo
  {pages} {389–399} (\bibinfo {year} {1963})}\BibitemShut {NoStop}%
\bibitem [{\citenamefont {Youngs}(2019)}]{Youngs2019}%
  \BibitemOpen
  \bibfield  {author} {\bibinfo {author} {\bibfnamefont {T.}~\bibnamefont
  {Youngs}},\ }\bibfield  {title} {\bibinfo {title} {{D}issolve: next
  generation software for the interrogation of total scattering data by
  empirical potential generation},\ }\href
  {https://doi.org/10.1080/00268976.2019.1651918} {\bibfield  {journal}
  {\bibinfo  {journal} {Mol. Phys.}\ }\textbf {\bibinfo {volume} {117}},\
  \bibinfo {pages} {3464–3477} (\bibinfo {year} {2019})}\BibitemShut
  {NoStop}%
\bibitem [{\citenamefont {B{\'e}e}(1988)}]{Bee1988}%
  \BibitemOpen
  \bibfield  {author} {\bibinfo {author} {\bibfnamefont {M.}~\bibnamefont
  {B{\'e}e}},\ }\href@noop {} {\emph {\bibinfo {title} {{Q}uasielastic neutron
  scattering, principles and applications in solid state chemistry, biology and
  materials science}}}\ (\bibinfo  {publisher} {CRC Press},\ \bibinfo {address}
  {London, UK},\ \bibinfo {year} {1988})\BibitemShut {NoStop}%
\bibitem [{\citenamefont {Coles}\ \emph {et~al.}(2023)\citenamefont {Coles},
  \citenamefont {Falkowski}, \citenamefont {Geddes}, \citenamefont {Pérez},
  \citenamefont {Booth}, \citenamefont {Squires}, \citenamefont {O’Rourke},
  \citenamefont {McColl}, \citenamefont {Goodwin}, \citenamefont {Cussen},
  \citenamefont {Clarke}, \citenamefont {Islam},\ and\ \citenamefont
  {Morgan}}]{Coles2023}%
  \BibitemOpen
  \bibfield  {author} {\bibinfo {author} {\bibfnamefont {S.~W.}\ \bibnamefont
  {Coles}}, \bibinfo {author} {\bibfnamefont {V.}~\bibnamefont {Falkowski}},
  \bibinfo {author} {\bibfnamefont {H.~S.}\ \bibnamefont {Geddes}}, \bibinfo
  {author} {\bibfnamefont {G.~E.}\ \bibnamefont {Pérez}}, \bibinfo {author}
  {\bibfnamefont {S.~G.}\ \bibnamefont {Booth}}, \bibinfo {author}
  {\bibfnamefont {A.~G.}\ \bibnamefont {Squires}}, \bibinfo {author}
  {\bibfnamefont {C.}~\bibnamefont {O’Rourke}}, \bibinfo {author}
  {\bibfnamefont {K.}~\bibnamefont {McColl}}, \bibinfo {author} {\bibfnamefont
  {A.~L.}\ \bibnamefont {Goodwin}}, \bibinfo {author} {\bibfnamefont {S.~A.}\
  \bibnamefont {Cussen}}, \bibinfo {author} {\bibfnamefont {S.~J.}\
  \bibnamefont {Clarke}}, \bibinfo {author} {\bibfnamefont {M.~S.}\
  \bibnamefont {Islam}},\ and\ \bibinfo {author} {\bibfnamefont {B.~J.}\
  \bibnamefont {Morgan}},\ }\bibfield  {title} {\bibinfo {title}
  {{A}nion-polarisation-directed short-range-order in antiperovskite
  \ce{Li2FeSO}},\ }\href {https://doi.org/10.1039/d2ta10037a} {\bibfield
  {journal} {\bibinfo  {journal} {J. Mater. Chem. A}\ }\textbf {\bibinfo
  {volume} {11}},\ \bibinfo {pages} {13016–13026} (\bibinfo {year}
  {2023})}\BibitemShut {NoStop}%
\bibitem [{\citenamefont {Legein}\ \emph {et~al.}(2024)\citenamefont {Legein},
  \citenamefont {Morgan}, \citenamefont {Squires}, \citenamefont {Body},
  \citenamefont {Li}, \citenamefont {Burbano}, \citenamefont {Salanne},
  \citenamefont {Charpentier}, \citenamefont {Borkiewicz},\ and\ \citenamefont
  {Dambournet}}]{Legein2024}%
  \BibitemOpen
  \bibfield  {author} {\bibinfo {author} {\bibfnamefont {C.}~\bibnamefont
  {Legein}}, \bibinfo {author} {\bibfnamefont {B.~J.}\ \bibnamefont {Morgan}},
  \bibinfo {author} {\bibfnamefont {A.~G.}\ \bibnamefont {Squires}}, \bibinfo
  {author} {\bibfnamefont {M.}~\bibnamefont {Body}}, \bibinfo {author}
  {\bibfnamefont {W.}~\bibnamefont {Li}}, \bibinfo {author} {\bibfnamefont
  {M.}~\bibnamefont {Burbano}}, \bibinfo {author} {\bibfnamefont
  {M.}~\bibnamefont {Salanne}}, \bibinfo {author} {\bibfnamefont
  {T.}~\bibnamefont {Charpentier}}, \bibinfo {author} {\bibfnamefont {O.~J.}\
  \bibnamefont {Borkiewicz}},\ and\ \bibinfo {author} {\bibfnamefont
  {D.}~\bibnamefont {Dambournet}},\ }\bibfield  {title} {\bibinfo {title}
  {{C}orrelated {A}nion {D}isorder in {H}eteroanionic {C}ubic \ce{TiOF2}},\
  }\href {https://doi.org/10.1021/jacs.4c06304} {\bibfield  {journal} {\bibinfo
   {journal} {J. Am. Chem. Soc.}\ }\textbf {\bibinfo {volume} {146}},\ \bibinfo
  {pages} {21889–21902} (\bibinfo {year} {2024})}\BibitemShut {NoStop}%
\bibitem [{\citenamefont {Armstrong}(2022)}]{Armstrong2022}%
  \BibitemOpen
  \bibfield  {author} {\bibinfo {author} {\bibfnamefont {J.}~\bibnamefont
  {Armstrong}},\ }\bibfield  {title} {\bibinfo {title} {{N}eutron spectroscopy
  as a method for classical force-field parameterization: {P}ast methods,
  present successes and future challenges},\ }\href
  {https://doi.org/10.1088/2399-6528/ac9728} {\bibfield  {journal} {\bibinfo
  {journal} {J. Phys. Comm.}\ }\textbf {\bibinfo {volume} {6}},\ \bibinfo
  {pages} {102002} (\bibinfo {year} {2022})}\BibitemShut {NoStop}%
\bibitem [{\citenamefont {Richardson}\ \emph {et~al.}(2026)\citenamefont
  {Richardson}, \citenamefont {McColl}, \citenamefont {Nilsen}, \citenamefont
  {Armstrong},\ and\ \citenamefont {McCluskey}}]{Richardson2026}%
  \BibitemOpen
  \bibfield  {author} {\bibinfo {author} {\bibfnamefont {H.}~\bibnamefont
  {Richardson}}, \bibinfo {author} {\bibfnamefont {K.}~\bibnamefont {McColl}},
  \bibinfo {author} {\bibfnamefont {G.~J.}\ \bibnamefont {Nilsen}}, \bibinfo
  {author} {\bibfnamefont {J.}~\bibnamefont {Armstrong}},\ and\ \bibinfo
  {author} {\bibfnamefont {A.~R.}\ \bibnamefont {McCluskey}},\ }\bibfield
  {title} {\bibinfo {title} {{L}ost in {T}ranslation: {S}imulation-{I}nformed
  {B}ayesian {I}nference {I}mproves {U}nderstanding of {M}olecular {M}otion
  {F}rom {N}eutron {S}cattering},\ }\href
  {https://doi.org/10.1021/acs.jpclett.6c01399} {\bibfield  {journal} {\bibinfo
   {journal} {J. Phys. Chem. Lett.}\ }\textbf {\bibinfo {volume} {17}},\
  \bibinfo {pages} {8411} (\bibinfo {year} {2026})}\BibitemShut {NoStop}%
\bibitem [{\citenamefont {McCluskey}\ \emph {et~al.}(2024)\citenamefont
  {McCluskey}, \citenamefont {Coles},\ and\ \citenamefont
  {Morgan}}]{McCluskey2024}%
  \BibitemOpen
  \bibfield  {author} {\bibinfo {author} {\bibfnamefont {A.~R.}\ \bibnamefont
  {McCluskey}}, \bibinfo {author} {\bibfnamefont {S.~W.}\ \bibnamefont
  {Coles}},\ and\ \bibinfo {author} {\bibfnamefont {B.~J.}\ \bibnamefont
  {Morgan}},\ }\bibfield  {title} {\bibinfo {title} {{A}ccurate {E}stimation of
  {D}iffusion {C}oefficients and their {U}ncertainties from {C}omputer
  {S}imulation},\ }\href {https://doi.org/10.1021/acs.jctc.4c01249} {\bibfield
  {journal} {\bibinfo  {journal} {J. Chem. Theory Comput.}\ }\textbf {\bibinfo
  {volume} {21}},\ \bibinfo {pages} {79–87} (\bibinfo {year}
  {2024})}\BibitemShut {NoStop}%
\bibitem [{\citenamefont {Fornies-Marquina}\ \emph {et~al.}(1997)\citenamefont
  {Fornies-Marquina}, \citenamefont {Letosa}, \citenamefont {Garcia-Gracia},\
  and\ \citenamefont {Artacho}}]{ForniesMarquina1997}%
  \BibitemOpen
  \bibfield  {author} {\bibinfo {author} {\bibfnamefont {J.~M.}\ \bibnamefont
  {Fornies-Marquina}}, \bibinfo {author} {\bibfnamefont {J.}~\bibnamefont
  {Letosa}}, \bibinfo {author} {\bibfnamefont {M.}~\bibnamefont
  {Garcia-Gracia}},\ and\ \bibinfo {author} {\bibfnamefont {J.~M.}\
  \bibnamefont {Artacho}},\ }\bibfield  {title} {\bibinfo {title} {{E}rror
  propagation for the transformation of time domain into frequency domain},\
  }\href {https://doi.org/10.1109/20.582534} {\bibfield  {journal} {\bibinfo
  {journal} {IEEE Trans. Magn.}\ }\textbf {\bibinfo {volume} {33}},\ \bibinfo
  {pages} {1456–1459} (\bibinfo {year} {1997})}\BibitemShut {NoStop}%
\bibitem [{\citenamefont {Zolzaya}\ and\ \citenamefont
  {McCluskey}(2026{\natexlab{a}})}]{FourierrorGithub}%
  \BibitemOpen
  \bibfield  {author} {\bibinfo {author} {\bibfnamefont {T.}~\bibnamefont
  {Zolzaya}}\ and\ \bibinfo {author} {\bibfnamefont {A.}~\bibnamefont
  {McCluskey}},\ }\href@noop {} {\bibinfo {title} {{fourierror-0.0.3}}},\
  \bibinfo {howpublished}
  {\href{https://github.com/scams-research/fourierror}{https://github.com/scams-research/fourierror}}
  (\bibinfo {year} {2026}{\natexlab{a}})\BibitemShut {NoStop}%
\bibitem [{\citenamefont {Ten~Eyck}(1973)}]{TenEyck1973}%
  \BibitemOpen
  \bibfield  {author} {\bibinfo {author} {\bibfnamefont {L.~F.}\ \bibnamefont
  {Ten~Eyck}},\ }\bibfield  {title} {\bibinfo {title} {{C}rystallographic fast
  {F}ourier transforms},\ }\href {https://doi.org/10.1107/s0567739473000458}
  {\bibfield  {journal} {\bibinfo  {journal} {Acta Crystallogr. A: Found.
  Adv.}\ }\textbf {\bibinfo {volume} {29}},\ \bibinfo {pages} {183–191}
  (\bibinfo {year} {1973})}\BibitemShut {NoStop}%
\bibitem [{\citenamefont {Kneller}\ \emph {et~al.}(1995)\citenamefont
  {Kneller}, \citenamefont {Keiner}, \citenamefont {Kneller},\ and\
  \citenamefont {Schiller}}]{Kneller1995}%
  \BibitemOpen
  \bibfield  {author} {\bibinfo {author} {\bibfnamefont {G.~R.}\ \bibnamefont
  {Kneller}}, \bibinfo {author} {\bibfnamefont {V.}~\bibnamefont {Keiner}},
  \bibinfo {author} {\bibfnamefont {M.}~\bibnamefont {Kneller}},\ and\ \bibinfo
  {author} {\bibfnamefont {M.}~\bibnamefont {Schiller}},\ }\bibfield  {title}
  {\bibinfo {title} {{nMOLDYN}: {A} program package for a neutron scattering
  oriented analysis of {M}olecular {D}ynamics simulations},\ }\href
  {https://doi.org/10.1016/0010-4655(95)00048-k} {\bibfield  {journal}
  {\bibinfo  {journal} {Comput. Phys. Commun.}\ }\textbf {\bibinfo {volume}
  {91}},\ \bibinfo {pages} {191–214} (\bibinfo {year} {1995})}\BibitemShut
  {NoStop}%
\bibitem [{\citenamefont {Cooley}\ and\ \citenamefont
  {Tukey}(1965)}]{Cooley1965}%
  \BibitemOpen
  \bibfield  {author} {\bibinfo {author} {\bibfnamefont {J.~W.}\ \bibnamefont
  {Cooley}}\ and\ \bibinfo {author} {\bibfnamefont {J.~W.}\ \bibnamefont
  {Tukey}},\ }\bibfield  {title} {\bibinfo {title} {{A}n algorithm for the
  machine calculation of complex {F}ourier series},\ }\href
  {https://doi.org/10.1090/s0025-5718-1965-0178586-1} {\bibfield  {journal}
  {\bibinfo  {journal} {Math. Comput.}\ }\textbf {\bibinfo {volume} {19}},\
  \bibinfo {pages} {297–301} (\bibinfo {year} {1965})}\BibitemShut {NoStop}%
\bibitem [{\citenamefont {Heideman}\ \emph {et~al.}(1984)\citenamefont
  {Heideman}, \citenamefont {Johnson},\ and\ \citenamefont
  {Burrus}}]{Heideman1984}%
  \BibitemOpen
  \bibfield  {author} {\bibinfo {author} {\bibfnamefont {M.}~\bibnamefont
  {Heideman}}, \bibinfo {author} {\bibfnamefont {D.}~\bibnamefont {Johnson}},\
  and\ \bibinfo {author} {\bibfnamefont {C.}~\bibnamefont {Burrus}},\
  }\bibfield  {title} {\bibinfo {title} {{G}auss and the history of the fast
  {F}ourier transform},\ }\href {https://doi.org/10.1109/massp.1984.1162257}
  {\bibfield  {journal} {\bibinfo  {journal} {IEEE ASSP Mag.}\ }\textbf
  {\bibinfo {volume} {1}},\ \bibinfo {pages} {14–21} (\bibinfo {year}
  {1984})}\BibitemShut {NoStop}%
\bibitem [{\citenamefont {Gentleman}\ and\ \citenamefont
  {Sande}(1966)}]{Gentleman1966}%
  \BibitemOpen
  \bibfield  {author} {\bibinfo {author} {\bibfnamefont {W.~M.}\ \bibnamefont
  {Gentleman}}\ and\ \bibinfo {author} {\bibfnamefont {G.}~\bibnamefont
  {Sande}},\ }\bibfield  {title} {\bibinfo {title} {Fast {F}ourier transforms:
  for fun and profit},\ }in\ \href {https://doi.org/10.1145/1464291.1464352}
  {\emph {\bibinfo {booktitle} {Proceedings of the AFIPS '66 Fall Joint
  Computer Conference}}},\ Vol.~\bibinfo {volume} {29}\ (\bibinfo  {publisher}
  {AFIPS Press},\ \bibinfo {address} {Washington, D.C.},\ \bibinfo {year}
  {1966})\ pp.\ \bibinfo {pages} {563--578}\BibitemShut {NoStop}%
\bibitem [{\citenamefont {Faber}\ and\ \citenamefont
  {Ziman}(1965)}]{Faber1965}%
  \BibitemOpen
  \bibfield  {author} {\bibinfo {author} {\bibfnamefont {T.~E.}\ \bibnamefont
  {Faber}}\ and\ \bibinfo {author} {\bibfnamefont {J.~M.}\ \bibnamefont
  {Ziman}},\ }\bibfield  {title} {\bibinfo {title} {{A} theory of the
  electrical properties of liquid metals: {III}. the resistivity of binary
  alloys},\ }\href {https://doi.org/10.1080/14786436508211931} {\bibfield
  {journal} {\bibinfo  {journal} {Phil. Mag.}\ }\textbf {\bibinfo {volume}
  {11}},\ \bibinfo {pages} {153–173} (\bibinfo {year} {1965})}\BibitemShut
  {NoStop}%
\bibitem [{\citenamefont {{Van Loan}}(1992)}]{VanLoan1992}%
  \BibitemOpen
  \bibfield  {author} {\bibinfo {author} {\bibfnamefont {C.~F.}\ \bibnamefont
  {{Van Loan}}},\ }\href@noop {} {\emph {\bibinfo {title} {{C}omputational
  {F}rameworks for the {F}ast {F}ourier {T}ransform}}},\ \bibinfo {series}
  {{F}rontiers in {A}pplied {M}athematics}, Vol.~\bibinfo {volume} {10}\
  (\bibinfo  {publisher} {SIAM},\ \bibinfo {address} {Philadelphia, US},\
  \bibinfo {year} {1992})\BibitemShut {NoStop}%
\bibitem [{\citenamefont {Zolzaya}\ and\ \citenamefont
  {McCluskey}(2026{\natexlab{b}})}]{Zolzaya2026}%
  \BibitemOpen
  \bibfield  {author} {\bibinfo {author} {\bibfnamefont {T.}~\bibnamefont
  {Zolzaya}}\ and\ \bibinfo {author} {\bibfnamefont {A.}~\bibnamefont
  {McCluskey}},\ }\href@noop {} {\bibinfo {title} {{fourier-prop-0.0.1}}},\
  \bibinfo {howpublished}
  {\href{https://github.com/scams-research/fourier-prop}{https://github.com/scams-research/fourier-prop}}
  (\bibinfo {year} {2026}{\natexlab{b}})\BibitemShut {NoStop}%
\end{thebibliography}%

\end{document}